# Fourier Transform Ultrasound Spectroscopy For the Determination of Wave Propagation Parameters

Barnana Pal

Saha Institute of Nuclear Physics,

1/AF, Bidhannagar, Kolkata-700064, India.

e-mail: barnana.pal@saha.ac.in

**Abstract:** The reported results for ultrasonic wave attenuation constant (α) in pure water show noticeable inconsistency in magnitude. A "Propagating-Wave" model analysis of the most popular pulse-echo technique indicates that this is a consequence of the inherent wave propagation characteristics in a bounded medium. In the present work Fourier Transform Ultrasound Spectroscopy (FTUS) is adopted to determine ultrasonic wave propagation parameters, the wave number (k) and attenuation constant (α) at 1MHz frequency in tri-distilled water at room temperature (25°C). Pulse-echo signals obtained under same experimental conditions regarding the exciting input signal and reflecting boundary wall of the water container for various lengths of water columns are captured. The Fast Fourier Transform (FFT) components of the echo signals are taken to compute k, α and r, the reflection constant at the boundary, using Oak Ridge and Oxford method. The results are compared with existing literature values.

**1. Introduction:** Pulse-echo method is the standard method widely used for measuring the velocity (v) and attenuation constant (α) of ultrasonic waves for characterizing solid and liquid samples including biological systems. In a recent work [1], it has been reported that, this method is not reliable for accurate measurement of intrinsic attenuation constant α. In fact, a common experience of experimentalists working with attenuation measurement in solid and liquid samples is that, it is very difficult to obtain reproducible results for α. An elaborative illustration in this point is the work by Martinez et.al. [2] on pure water where both pulse-echo and through-transmission methods are used to measure α. It is observed that at relatively small distances from the transducer, α varies widely, while at large distances it is possible to obtain an average value of α. Moreover, the values of α obtained from pulse-echo and through transmission methods differ significantly and do not agree with other values reported in the literature [3-6]. The propagating-wave analysis of pulse-echo method [1] gives a satisfactory explanation for the variation. It shows that the echo amplitudes depend on α, the reflection coefficient r at the boundaries and the sample length l in a complicated way. As a consequence, the attenuation measured from the echo heights is an effective attenuation $α_e$, different from the intrinsic attenuation α of the propagating medium. The dependence of $α_e$ on r was reported [7] in solid samples where the effect is more prominent due to the presence of the couplant used for bonding the transducer to the sample. We show that the intrinsic attenuation α along with r can be computed from the frequency spectrum of the pulse-echo signal using Oak Ridge and Oxford parameter fitting method [8]. This method has been used to determine α in tri-distilled water at room temperature (25°C) at the transducer frequency 1 MHz. The experimental method is described in sec. 2, sec.3 presents the results and discussion and sec. 4 gives the conclusion.

## 2. Experimental procedure and analysis:

Ultran HE900 rf burst generator is used to generate the input exciting signal. The carrier frequency (f) of the rf pulse is 1 MHz, pulse width is ~3 μs, pulse repetition time is ~100 ms, peak-to-peak height of the pulse is ~250V. This carrier frequency pulse train may be considered as the superposition of a large number of continuous waves with frequencies and amplitudes determined by its Fourier components [9]. According to the "propagating wave" model description, a limited number of these components, having frequencies within the bandwidth of the transducer loaded with the sample, will be converted to mechanical waves by the transducer and propagate through the medium under investigation[10,11]. For the present experiment with tri-distilled water at room temperature maintained at $25^0C$, water is taken in cylindrical containers of same diameter and of different lengths (l). A circular reflector plate with plane surface is kept at the bottom to reflect ultrasonic wave so that reflection coefficient (r) remains same at each measurement. The distance (l) from the transducer face to the reflector plate is measured using slide calipers. The component waves of the input signal lying within the bandwidth of loaded transducer will suffer successive multiple reflections at the water-reflector interface at the bottom of the container and water-transducer interface at the transducer end which are kept parallel to each other. At each reflection, the amplitude of the wave will be modified by a factor r with associated phase reversal since in this particular case the reflection is from the boundary of higher mechanical impedance. All of these reflected waves will superpose inside the sample and in the steady state the displacement wave at any position x at time t may be written as (1),

$$U(x,t) = \sum_\omega A_\omega e^{i\omega t} \quad (1)$$

The summation is over all the relevant frequency components.

The wave amplitude $A_\omega$ is given by,

$$A_\omega = \frac{a[e^{-ik'x} - re^{-ik'(2l-x)}]}{1-rr'e^{-2ik'l}} \quad (2)$$

Here, $a$ is the amplitude of the component wave of frequency ω at source point, $k$ is the propagation constant (= 2πf/v), $r$ is the reflection co-efficient at water-reflector interface, $r'$ is the reflection co-efficient at water-transducer interface, $k' = k - i\alpha$ and x is the position measured from the transducer face. For simplicity, we take $r = r'$. In the present experiment, since a single transducer is used as the transmitter as well as the receiver, the response observed in the oscilloscope will be proportional to the wave displacement at x=0, i.e. $U(0,t)$.

The echo trains for n different sample lengths, $l_1$, $l_2$, ......., $l_n$, are captured using digital oscilloscope DL 1640 and stored in PC for further analysis. From the average separation Δt between successive echoes, ultrasound velocity v is determined (v=2$l_n$/Δt) and $k$ is calculated. From exponential fitting of the echo heights, effective attenuation $\alpha_e$ is determined. To get the intrinsic attenuation constant α the following procedure is adopted. The Fast Fourier Transform (FFT) of the echo trains obtained for n lengths are determined. For each frequency component, n wave amplitudes are obtained for n different sample lengths. Oak Ridge and Oxford method for parameter fitting [8] is used to fit these n amplitudes according to relation (2) by adjusting the parameters k, r, α and $a$. The computational method requires input guess values for k, r, α and $a$. Input k is obtained from experimentally measured v, input α is the lowest value of $\alpha_e$ obtained for large l, input r is calculated using the relation r = $(\rho_1v_1 - \rho_2v_2)/(\rho_1v_1 + \rho_2v_2)$, ρ being the density, with suffixes 1 and 2 designating the reflector material and water respectively, and input $a$ is chosen arbitrarily. The experiment is repeated using steel, copper, brass, lead and glass reflectors.

## 3. Results and discussion

The velocities (v) and effective attenuation constants ($\alpha_e$) are determined for various lengths (l) of water columns in pulse-echo experiment with five different reflecting surfaces. The values for v are close to each other within experimental error. Average v is determined to be $1.4775 \times 10^5$ cm.sec$^{-1}$. Figure-1shows the dependence of $\alpha_e$ on l for glass reflector. We see that attenuation values are widely different particularly for small values of l. Similar variation in the attenuation constant has been reported by R Martinez and co-workers [2]. Their work shows clearly that at short distances from the transmitting transducer, the attenuation values measured in pulse-echo method show wide variation. At long distances however, the results are consistent and it is possible to get an average value of 0.04417 np/cm for the attenuation constant though it is not in agreement with other reported values [3-6]. Their measurement in through-transmission method uses two transducers aligned face-to-face, parallel to each other. By this the water medium in between behaves effectively like a bounded medium as in pulse-echo method and no better solution is obtained. The attenuation value obtained in this method is 0.04654 np/cm.

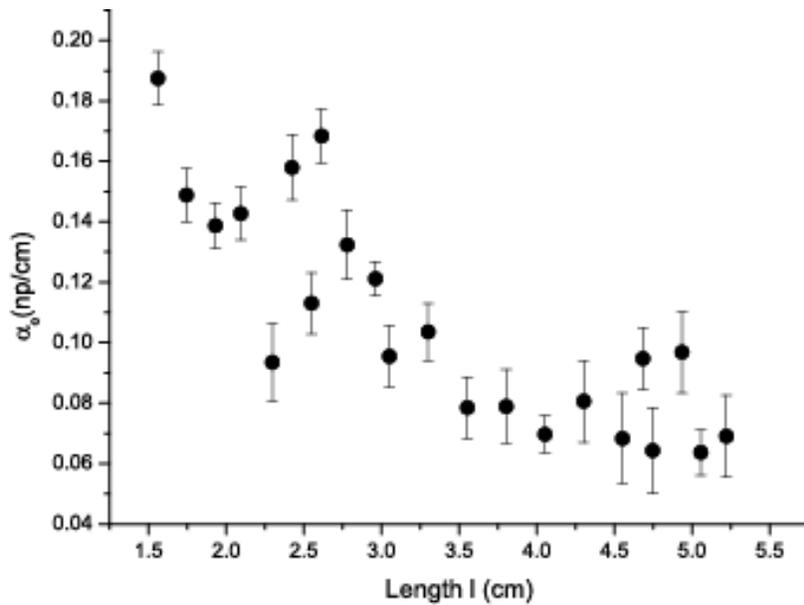

Figure 1: Variation of effective attenuation $\alpha_e$ with length l of water column using glass reflector

Measurements with other four reflectors, viz., steel, copper, brass and lead show similar nature of variation of $\alpha_e$ with l and this is illustrated in figure 2. No significant difference is noticed due to the differences in r because of the fact that fluctuation due to change in l is more prominent than the effect caused by the change in r. For longer l and smaller r the echo signals are weak and measurement of $\alpha_e$ is more erroneous. The observed nature of fig. 2 is also consistent with the numerical study presented in ref [1].

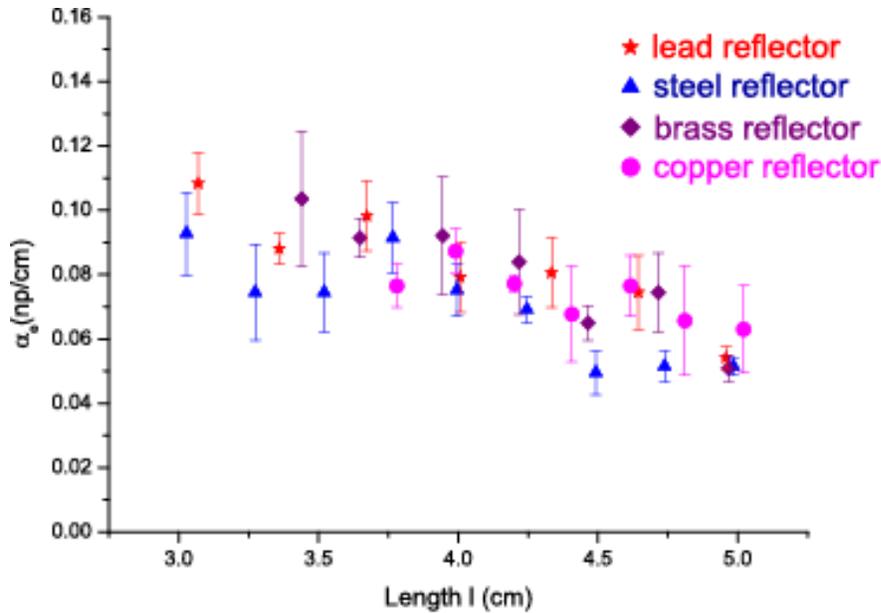

Figure 2: Variation of effective attenuation $\alpha_e$ with length l of water column for lead, steel, brass and copper reflector.

To determine the intrinsic attenuation constant α, FFT components are computed for pulse-echo signals captured for seven different column lengths ($l_n$, n=1,2,.........7) under same exciting conditions. The exciting input is pulsed rf signal of carrier frequency 1 MHz, peak-peak pulse height of ~300 V, pulse width of ~ 3 μs and pulse repetition time ~ 2.45 ms. Table-I presents the liquid column lengths for different reflectors used to get pulse-echo signals. Figure3 shows representative echo signal captured with glass reflector for (a) $l_1$= 5.056 cm and (b) $l_7$= 3.552 cm and figure 4(a) and (b) show their fft components respectively. In fig 4 unit frequency corresponds to 100 MHz and relative amplitude is in arbitrary unit. The frequency range where the fft amplitudes are relatively high is shown in fig 4. The peak region is considered for computational parameter fitting. The error in experimental fft amplitudes is assumed to be within 10%. Best fit parameters with minimum $\chi^2$ error are determined using Oak Ridge and Oxford method [8].

**Table-I: Water column lengths.**

| Reflector material | Water column lengths in cm | | | | | | |
|---|---|---|---|---|---|---|---|
| | $l_1$ | $l_2$ | $l_3$ | $l_4$ | $l_5$ | $l_6$ | $l_7$ |
| Steel | 4.984 | 4.738 | 4.492 | 4.246 | 3.996 | 3.766 | 3.520 |
| Copper | 5.022 | 4.810 | 4.616 | 4.406 | 4.200 | 3.990 | 3.780 |
| Brass | 4.968 | 4.716 | 4.464 | 4.218 | 3.944 | 3.694 | 3.440 |
| Lead | 4.958 | 4.644 | 4.334 | 4.008 | 3.674 | 3.360 | 3.070 |
| Glass | 5.056 | 4.804 | 4.552 | 4.302 | 4.052 | 3.802 | 3.552 |

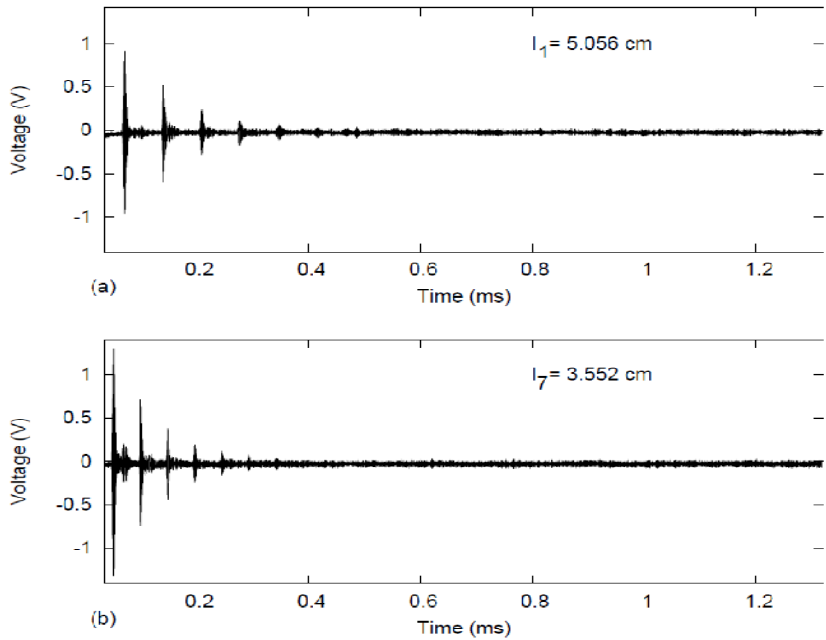

Figure 3: Echo signal using glass reflector for (a) $l_1$ = 5.056 cm and (b) $l_7$ = 3.552 cm

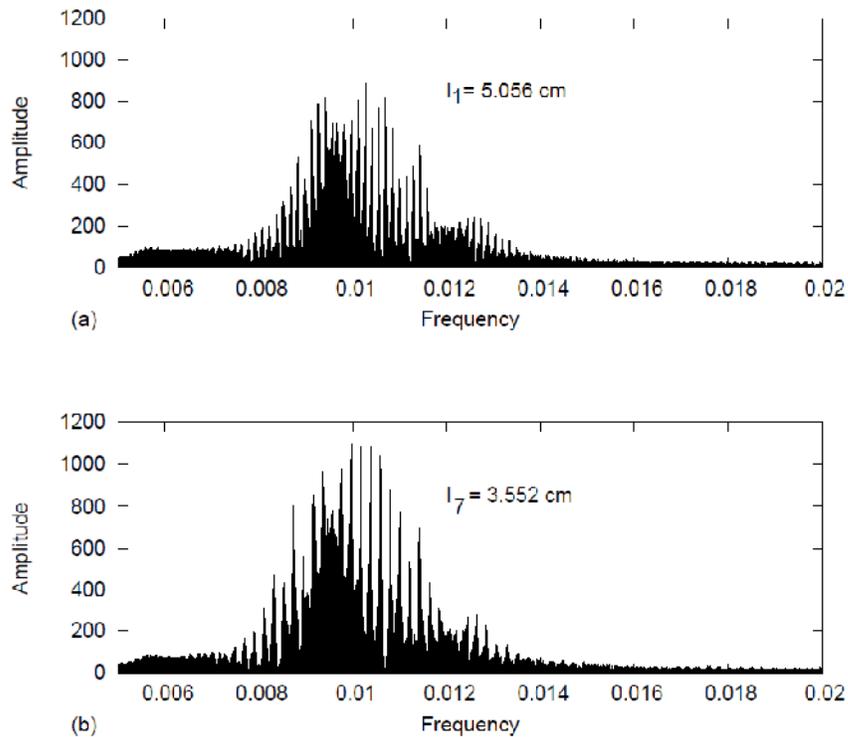

Figure 4: Frequency spectrum for echo signal obtained with glass reflector showing peak amplitudes for (a) $l_1$ = 5.056 cm and (b) $l_7$ = 3.552 cm

These parameters along with their input values are presented in Table-II. The average k thus obtained is 42.74 ± 0.03 cm$^{-1}$ giving v = (1.4693 ± 0.0011) × 10$^5$ cm/sec and average α is 0.0435 ± 0.0013 np/cm. This value of α is more accurate and consistent with the measurement by R Martinez et. al. [2].

**Table-II: Input parameters and best fit output parameters**

| Reflector material | Input parameters | | | Best fit output parameters | | | | |
|---|---|---|---|---|---|---|---|---|
| | k (cm$^{-1}$) (experimental) | r (calculated) | $α_e$ (np.cm$^{-1}$) | k (cm$^{-1}$) | Av k (cm$^{-1}$) | r | α (np.cm$^{-1}$) | Av α (np.cm$^{-1}$) |
| Steel | 42.72 | 0.94 | 0.0497 | 42.76 | 42.74±0.03 | 0.92 | 0.0439 | 0.0435±0.0013 |
| Copper | | 0.93 | 0.0619 | 42.72 | | 0.91 | 0.0422 | |
| Brass | | 0.93 | 0.0510 | 42.77 | | 0.91 | 0.0455 | |
| Lead | | 0.89 | 0.0517 | 42.69 | | 0.87 | 0.0418 | |
| Glass | | 0.81 | 0.0639 | 42.76 | | 0.80 | 0.0440 | |

### 4. Conclusion

We have measured v and $α_e$ in tri-distilled water as functions of water column length l, using five different reflectors with different reflection constant r, in the pulse-echo method at 1 MHz transducer frequency at room temperature 25$^0$C. The observed nature of variation of $α_e$ is well in agreement with our previous numerical analysis [1] as well as those reported by R. Martinez et. al. [2]. The pulse-echo signals are further used to determine the intrinsic attenuation constant α by Fourier Transform Ultrasound Spectroscopy (FTUS). This method uses Oak Ridge and Oxford method [8] of standard $χ^2$ parameter fitting on the fft amplitudes of the pulse-echo signals with theoretical amplitudes as calculated using propagating model, over the frequency range where these amplitudes are considerably high. The attenuation constant α thus obtained, though differs widely from other literature values due to reasons explained in ref. [1], is more accurate and agrees with the value of R. Martinez et. al. [2].

**Acknowledgement**:

The author is grateful to Sankari Chakraborty, Papia mondal, Anish Karmahapatro and Arindam Chakrabarti for technical and other assistances during experiment and data analysis.